\documentclass[12pt]{article}
\usepackage{amsmath,amssymb,epsfig}
\usepackage{cancel}
\usepackage{color}
\usepackage{bm}
\usepackage{braket}

\makeatletter

\@addtoreset{equation}{section}
\makeatother

\newcommand{\im}{{\rm Im\,}}

\begin{document}

\title{\vbox{
\baselineskip 14pt
\hfill \hbox{\normalsize WU-HEP-17-04, EPHOU-17-003, KUNS-2667}
} \vskip 1.7cm
\bf K\"ahler moduli stabilization 
in semi-realistic magnetized orbifold models
\vskip 0.5cm
}
\author{
Hiroyuki~Abe$^{1}$,
Tatsuo~Kobayashi$^{2}$, \
Keigo~Sumita$^{1}$, \ and \
Shohei~Uemura$^{3}$
\\*[20pt]
{\it \normalsize 
${}^{1}$Department of Physics, Waseda University, 
Tokyo 169-8555, Japan}
\\
{\it \normalsize 
${}^{2}$Department of Physics, Hokkaido University, 
Sapporo, 060-0810 Japan}
\\
{\it \normalsize 
${}^{3}$Department of Physics, Kyoto University, 
Kyoto 606-8502, Japan}
\\*[50pt]}

\date{
\centerline{\small \bf Abstract}
\begin{minipage}{0.9\linewidth}
\medskip 
\medskip 
\small
We study K\"ahler moduli stabilizations 
in semi-realistic magnetized D-brane models based on 
$ Z_2\times Z_2'$ toroidal orbifolds. 
In type IIB compactifications, 3-form fluxes can 
stabilize the dilaton and complex structure moduli fields, 
but there remain some massless closed string moduli fields, 
K\"ahler moduli. 
The magnetic fluxes generate Fayet-Iliopoulos terms, which can fix ratios of 
K\"ahler moduli.
On top of that, we consider D-brane instanton effects to stabilize them 
in concrete D-brane models and investigate 
the brane configurations to confirm that 
the moduli fields can be stabilized successfully. 
In this paper, 
we treat two types of D-brane models. One is based on D9-brane systems 
respecting the Pati-Salam model. The other is realized 
in a D7-brane system breaking the Pati-Salam gauge group. 
We find suitable configurations where 
the D-brane instantons can stabilize the moduli fields 
within both types of D-brane models, 
explaining an origin of a small constant term of the superpotential 
which is a key ingredient for successful moduli stabilizations. 
\end{minipage}
}

\newpage
\begin{titlepage}
\maketitle
\thispagestyle{empty}
\clearpage
\tableofcontents
\thispagestyle{empty}
\end{titlepage}

\renewcommand{\thefootnote}{\arabic{footnote}}

\section{Introduction}
Superstring theories are expected for an ultimate unified theory of 
particle physics including gravitational interactions. 
One of their remarkable features is that 
superstring theories are defined in ten-dimensional (10D) spacetime and 
predict the presence of extra dimensions of space 
for theoretical consistencies.
We usually consider that 
the extra six-dimensional (6D) space is compactified 
in order to describe our universe.

In such string compactifications, one of challenging tasks is 
to realize a chiral spectrum 
in their four-dimensional (4D) effective theories, 
because they must be consistent with the standard model (SM) or 
some extensions such as 
minimal supersymmetric standard model (MSSM). 
For the purpose, 
D-brane models are attractive because they can lead to 
various gauge groups with generations of chiral fermions
\cite{Bachas:1995ik,Berkooz:1996km,Blumenhagen:2000wh,
Aldazabal:2000dg,Angelantonj:2000hi}, and 
several D-brane models were proposed 
realizing suitable 4D chiral spectra as zero-modes of 
open strings on intersecting D-branes~\cite{Ibanez:2001nd,Cremades:2002qm,
Honecker:2004kb, Cvetic:2001nr}. 
For the last decade, similar model building was 
actively attempted in their T-dual picture, that is, 
in the framework of IIB strings with magnetized D-branes, 
and it was found that viable three-generation models can be obtained 
\cite{Cremades:2004wa,Abe:2008sx,Abe:2015yva}. 
In particular, in a concrete model proposed in Ref.~\cite{Abe:2012fj}, 
a semi-realistic flavor structure of the quarks and the leptons 
including their hierarchical masses and mixing angles was obtained, 
and furthermore, a spectrum of the supersymmetric particles and 
the Higgs bosons was calculated to verify its consistency 
with experimental results.

Another one of the key issues in the string compactifications is 
stabilization of moduli fields which are massless scalar modes 
originating from extra components of the higher-dimensional gravitational 
fields and n-form fields. 
Moduli stabilization is necessary to stabilize the extra compact space, 
and that is also significant in particle and 
cosmological phenomenologies. 
In these decades, several moduli stabilization mechanisms are proposed 
in the framework of superstring theories. 
We will discuss the moduli stabilization, concentrating 
on type IIB compactifications in this paper 
to associate them with magnetized D-brane models 
(Moduli stabilizations with the magnetic fluxes were discussed 
in Refs.~\cite{Antoniadis:2004pp,Antoniadis:2005nu,Antoniadis:2006eu}). 
We find three types of dynamical variables to be stabilized, 
dilaton field, complex structure moduli and K\"ahler moduli fields. 
Basically, in IIB string theories, 
we can introduce nontrivial fluxes for 3-form field strengths 
to stabilize the dilaton and complex structure moduli fields~\cite{Gukov:1999ya,Giddings:2001yu}. 
In the presence of the 3-form fluxes turned on, however, 
the potential for the K\"ahler moduli keeps flat at the tree level, 
and there remain some flat directions 
even when $\alpha'$-corrections and string 1-loop corrections 
are taken into account. 
We usually expect that 
those flat directions of K\"ahler moduli fields are 
stabilized by nonperturbative effects somehow.

In D-brane models, one of computable nonperturbative effects is  
D-brane instantons~\cite{Blumenhagen:2006xt, 
Ibanez:2006da, Ibanez:2007rs, Cvetic:2007ku,Blumenhagen:2009qh}, 
which we call Euclidean-branes (E-branes) in the present paper. 
That is D-branes localized at a point on 4D Minkowski spacetime 
but has a nonzero volume on the extra compact space. 
Thus, they are possible 
to yield a superpotential for the K\"ahler moduli and the dilaton field. 
Besides that, gaugino condensations of hidden D-branes 
are also  computable nonperturbative effects 
to yield the superpotential 
of the moduli fields, but we will focus on the former one in this paper. 
% can induce the superpotential for K\"ahler moduli, since the tree level gauge
% coupling of Yang-Mills theory in the D-brane world volume 
%is given by the inverse of the D-brane volume.
%Considering above nonperturbative effects, we can stabilize K\"ahler moduli.
%The KKLT model and the LARGE volume scenario are 
%representative two examples\c
%ite{Kachru:2003aw,Balasubramanian:2005zx}. 

In most of previous works\footnote{
There are several studies of moduli stabilizations in D-brane models, see 
Refs.~\cite{Blumenhagen:2005tn,Blumenhagen:2007sm,
Cicoli:2011qg,Cicoli:2012bi}}, 
D-brane model building for the visible sector and 
the moduli stabilization is discussed independently from each other. 
Such a scenario can be justified under the situation that 
the visible sector is irrelevant to the sector to stabilize moduli. %e.g. E-branes, 
% and 
%the scale of stabilization dynamics is higher than the weak scale.
For example, if the SM sector is localized at a certain point on the 6D compact space 
and the dynamics to stabilize moduli originates from the sector on cycles far away from 
the SM-localized point, those would be independent.
However, if the SM sector and the moduli-stabilizing sector occupy 
at a similar place in the 6D compact space, they would affect each other.
%That is indeed justified by their energy scales as follows.
%Naively, a typical mass scale of the moduli fields must be 
%comparable to the string scale and that is 
%far away from the electroweak scale, except for 
%particular scenarios~\cite{Balasubramanian:2005zx}. 
%Thus we can replace the moduli fields by their 
%vacuum expectation values (VEVs) in the low-energy effective field theories, 
%and they are treated as free-parameters there. 
%except a few works 
%(e.g.\cite{Blumenhagen:2007sm,Cicoli:2011qg,Cicoli:2012bi}) 
Indeed, it is not trivial that 
the instanton effect yields a superpotential suitable for 
the moduli stabilization such as $W \sim A e^{-aT}$, where $T$ is the modulus,  
and that in practice depends on configurations of D-branes 
for the visible sector. 
This is due to the fact that 
one needs to integrate over the instanton zero-modes 
to obtain nonperturbative superpotentials. 
We can realize the superpotential successfully when 
there is only a single E-brane wrapping $O(1)$-cycles without D-branes. 
On the other hand, 
in association with D-branes, 
there appear open string zero-modes between the E-branes and the D-branes.
When they can not be soaked up by fermionic integration, the nonperturbative superpotential vanishes.
Furthermore, even if zero-modes are successfully soaked up, the superpotential including matter fields can be induced 
as $W \sim (\Phi_1 \Phi_2 \cdots ) e^{-aT}$, but not the pure moduli term $W \sim A e^{-aT}$.
Such moduli-dependent terms with matter fields would be important to realize the right-handed Majorana neutrino masses and 
$\mu$-terms of the Higgs fields in MSSM 
\cite{Blumenhagen:2006xt,Ibanez:2006da,Ibanez:2007rs,Cvetic:2007ku,Blumenhagen:2009qh,Kobayashi:2015siy,Kobayashi:2016ovu}.
However, such moduli-dependent terms with matter fields are not suitable for moduli stabilizations. 
% in the superpotential
%generate a scalar potential including matter fields, and it would not stabilize all the scalar fields including moduli at once.
%Hence, nonperturbative terms including only moduli such as $W \sim A e^{-aT}$ are favorable.
%and the D-branes, and then, 
%the nonperturbative superpotential will vanish because of 
%the integration over such extra fermionic zero-modes. 
We are thus required to study distributions of the zero-modes 
for each brane configuration and confirm that 
no harmful fermionic zero-modes remain 
to incorporate the moduli stabilizations with the D-brane models.

In this paper, we study 
the moduli stabilization due to the E-branes 
in association with concrete magnetized 
D-brane models for the visible sector 
in type IIB orientifolds. 
We assume the 3-form fluxes to stabilize 
the dilaton and the complex structure 
moduli fields preserving supersymmetry (SUSY), which allow us to 
concentrate on the K\"ahler moduli stabilization\footnote{
Strictly speaking, we assume that 
the 3-form fluxes do not change the toroidal geometry so much, 
and blow-up moduli fields are set to zero.}. 
In those models, 
we will also turn on the ``magnetic'' fluxes for 
worldvolume gauge field strength of the D-branes 
in order to realize the flavor structure of the SM. 
These magnetic fluxes classically produce 
moduli depending Fayet-Iliopoulos (FI) terms. 
We will find supersymmetric vacua with a certain ratio of 
the VEVs of the moduli fields, 
that means the D-term potential can stabilize the K\"ahler 
moduli fields except for one flat direction. 
In order to stabilize the flat direction, 
we introduce E-branes and investigate the zero-mode structure 
in the D-brane models .

This paper is organized as follows. 
In section \ref{sec:d9}, we first review the 
magnetized $T^6/Z_2\times Z_2'$ orbifolds in 10D SYM theories, which correspond to 
the low-energy effective field theory of  D9-brane systems, 
which explains an essence of magnetized orbifold models. 
Consequently, we propose several concrete models based on 
the Pati-Salam gauge group. 
Two types of E-branes are possible to give 
stable brane configurations in association with D9-branes. 
In the rest of the section, we study both the instanton effects 
to find several brane configurations with which 
the instanton effects work successfully and 
the induced superpotential stabilizes the moduli field. 
In section \ref{sec:D7-brane}, 
we perform a similar analysis with D7-brane models of the visible sector 
where the Pati-Salam gauge group is broken by the magnetic fluxes 
to realize a more realistic spectrum. 
%The magnetized D-branes are T-dual to intersecting D-branes and 
%we can also investigate the structure of the zero-modes 
%in the T-dual picture, which we will discuss 
%in section \ref{sec:T-dual}. 
Section~\ref{sec:con} is devoted to 
conclusions and discussions. 
In Appendix A, we discuss the zero-mode structure in T-dual picture.

\section{D9-brane models} 
\label{sec:d9}
We  study mixed configurations of magnetized D-branes 
and E-branes to 
construct models with all the moduli fields stabilized. 
In this section, we focus on 
Pati-Salam models based on a stack of eight D9-branes as 
the SM sector. These are the simplest but semi-realistic 
magnetized orbifold models. First we briefly review the 10D SYM theories 
compactified on magnetized orbifold which are low-energy effective field theories of  magnetized D9-branes. 
In the theories, 
we can find several semi-realistic models based 
on the Pati-Salam gauge group. 
Finally, we will investigate E-brane's effects in the D9-brane systems,
which generate nonperturbative superpotential and stabilize 
the moduli fields.
Note that any configuration of E-branes can appear and we have to 
take into account all the possible E-branes.
Some of them have no effects in low-energy effective field theory, but 
a certain E-brane can have nonperturbative moduli terms such as $W \sim Ae^{-aT}$.
We are interested in such E-brane effects.

\subsection{Review of magnetized orbifolds in 10D SYM theories}
We give an overview on magnetized orbifold in 10D SYM theories. 
In this paper, we consider three 2-tori, $T^2\times T^2\times T^2$, 
as an extra compact space, denoting their coordinates by 
$z_i$ and $\bar{z}_i$ ($i=1,2,3$). 
The 10D SYM theories can be described in the formulation of 
4D $\mathcal N=1$ superspace, 
focusing on a 4D $\mathcal N=1$ SUSY out of full $\mathcal N=4$ SUSY of 
the 10D SYM theories~\cite{Marcus:1983wb}. 
This was developed in compactifications of $T^2\times T^2\times T^2$ 
with magnetic fluxes in Ref~\cite{Abe:2012ya}. 
10D SYM theories consist of 10D vector and Majorana-Weyl spinor 
fields, which are decomposed into 4D vector, complex scalar and Weyl 
spinor fields. These 4D fields form 4D $\mathcal N=1$ supermultiplets. 
As a result, field contents of the theories are expressed by 
a vector superfield $V$ and three chiral superfields $\Phi_i$. 
Note that they are in adjoint representations of 
gauge symmetry of the SYM theories. 
In the following, we consider $U(N)$ SYM theories as 
effective field theories of one stack of $N$ D9-branes.

We introduce Abelian magnetic fluxes in the $U(N)$ theories, 
which are parametrized by $N\times N$ diagonal matrices as 
\begin{equation*}
M^{(i)}={\rm diag}\,(m_1^{(i)},m_2^{(i)},\ldots,m_N^{(i)}), 
\end{equation*}
where $i$ runs over $1,2,3$ corresponding to three $T^2$. 
When $m_n^{(i)}$ takes nondegenerate values $U(N)$ gauge group is 
broken down.
For example,  suppose the simplest case as follows, 
\begin{equation}
M^{(i)}={\rm diag}\,(m_a^{(i)},\ldots,m_a^{(i)},m_b^{(i)},
\ldots,m_b^{(i)}), \label{eq:simp}
\end{equation}
where $m_a^{(i)}\neq m_b^{(i)}$. 
Then, these magnetic fluxes break the gauge group as 
$U(N)\rightarrow U(N_a)\times U(N_b)$. 
In this gauge symmetry breaking, we  express the superfields as 
\begin{equation}
\Phi_i\rightarrow
\begin{pmatrix}
\Phi_i^{aa}&\Phi_i^{ab}\\
\Phi_i^{ba}&\Phi_i^{bb}
\end{pmatrix}\label{eq:decomp}, 
\end{equation}
where diagonal and off-diagonal entries are in adjoint and bifundamental 
representations of the unbroken gauge group $U(N_a)\times U(N_b)$, respectively. 
On this magnetized background, zero-mode equations for $\Phi_j^{ab}$ 
on the $i$-th $T^2$ are given by 
\begin{eqnarray}
\left[\bar\partial_{\bar i}+\frac{\pi}{2\im\tau_i}
(m_a^{(i)}-m_b^{(i)})z_i\right]\Phi_j^{ab}&=&0\qquad({\rm for}~~i=j),
\label{eq:zeroii}\\
\left[\partial_{i}-\frac{\pi}{2\im\tau_i}
(m_a^{(i)}-m_b^{(i)})\bar z_{\bar i}\right]\Phi_j^{ab}&=&0
\qquad({\rm for}~~i\neq j), \label{eq:zeroij}
\end{eqnarray}
where $\tau_i$ is a complex structure of the $i$-th $T^2$. 
For $i=j$, that has $m_a^{(i)}-m_b^{(i)}$ degenerate zero-modes 
when $m_a^{(i)}-m_b^{(i)}$ is positive, 
while its conjugate one $\Phi_j^{ba}$ has no zero-modes because of 
$m_b^{(i)}-m_a^{(i)}<0$. 
Thus, the magnetic fluxes produce 
generations of chiral fermions in 4D effective theories. 
This is almost the same for $i\neq j$, except for 
a relative sign in Eq.~(\ref{eq:zeroij}), and 
$|m_a^{(i)}-m_b^{(i)}|$ degenerate zero-modes are produced for 
$\Phi_j^{ab}$ when $m_a^{(i)}-m_b^{(i)}$ is negative.

Next we study $ Z_2$ orbifolding in this magnetized SYM theories. 
Let us consider $ Z_2$ orbifolding which acts on the first and 
the second $T^2$, that is, 
\begin{equation*}
(z_1,z_2,z_3)\rightarrow (-z_1,-z_2,z_3). 
\end{equation*}
On this orbifold, the superfields have to transform as 
\begin{eqnarray}
V(z_1,z_2,z_3) &\rightarrow& +PV(-z_1, -z_2,  z_3)P^{-1},\nonumber\\
\Phi_1(z_1,z_2,z_3) &\rightarrow& - P\Phi_1(-z_1, -z_2, z_3)P^{-1},
\nonumber\\
\Phi_2(z_1,z_2,z_3) &\rightarrow& -P\Phi_2(-z_1, -z_2, z_3) P^{-1},
\nonumber\\
\Phi_3(z_1,z_2,z_3) &\rightarrow& + P\Phi_3(-z_1, -z_2, z_3)P^{-1},
\label{eq:Z_2transform}
\end{eqnarray}
where projection operator $P$ is an $N\times N$ 
matrix satisfying $P^2=\bm 1$. 
In accordance with these transformation laws, each entry of 
Eq.~(\ref{eq:decomp}) is assigned into either $ Z_2$ even or odd mode. 
This $ Z_2$ projection reduces the number of 
the degenerate zero-modes induced by the magnetic fluxes, 
as shown in Table~\ref{tb:numzero}~\cite{Abe:2008fi}. 
We can also introduce discrete Wilson lines, 
and the number of zero-modes depends on values of discrete Wilson lines \cite{Abe:2013bca}.
Here, for simplicity, we restrict ourselves to models with vanishing Wilson lines.

\begin{table}[t]
\center
\begin{tabular}{ccccccccc}
 $|M|$&$0$& $1$  &$2$  &$3$  &$4$  &$5$  &$2n$&$2n+1$ \\\hline
Even& $1$  &$1$  &$2$  &$2$  &$3$  &$3$ &$n+1$&$n+1$\\
Odd& $0$  &$0$  &$0$  &$1$  &$1$  &$2$  &$n-1$&$n$\\
\end{tabular}
\caption{The number of active zero-modes on 
the magnetized orbifold is shown, where $M$ represents an 
effective magnetic flux (That corresponds to $m_a^{(i)}-m_b^{(i)}$ in 
Eqs.~(\ref{eq:zeroii}) and (\ref{eq:zeroij}).).} 
\label{tb:numzero}
\end{table}

It is most important that the Abelian magnetic fluxes 
generically induce the FI-term 
for trivial $U(1)$ parts of 
unbroken gauge subgroups. For instance, 
in the case of Eq.~(\ref{eq:simp}), 
there appear the FI-terms with the following 
parameters in diagonal parts $U(1)_a \times U(1)_b$ of 
$U(N_a)$ and $U(N_b)$, 
\begin{eqnarray*}
\xi_a&=&\frac{1}{\mathcal{A}^{(1)}}m_a^{(1)}+
\frac{1}{\mathcal{A}^{(2)}}m_a^{(2)} 
+\frac{1}{\mathcal{A}^{(3)}}m_a^{(3)},
\label{eq:fia}\\
\xi_b&=&\frac{1}{\mathcal{A}^{(1)}}m_b^{(1)}+
\frac{1}{\mathcal{A}^{(2)}}m_b^{(2)} 
+\frac{1}{\mathcal{A}^{(3)}}m_b^{(3)},
\label{eq:fib}
\end{eqnarray*}
where $\mathcal{A}^{(i)}$ is the area of the $i$-th $T^2$. 
When setting $\mathcal{A}^{(i)}$ for $\xi_a$ and $\xi_b$ to vanish, 
we can find a supersymmetric vacuum with unbroken $U(N_a)$ and $U(N_b)$ 
gauge symmetries\footnote{
Magnetized supersymmetric 
vacua with broken $U(N_a)$ and $U(N_b)$ can also exist 
when charged fields develop their nonvanishing VEV in D-flat directions. 
This was discussed in Ref.~\cite{Abe:2016jsb}}. 
This means that some of the K\"ahler moduli fields 
are stabilized by the D-term potential at the supersymmetric vacuum. 
%Then, the $U(1)_a \times U(1)_b$ gauge bosons become massive.
In the present case, only the ratios of $\mathcal{A}^{(i)}$ 
are completely determined 
unless $m_a^{(1)}=m_a^{(2)}=m_a^{(3)}=0$ and/or 
$m_b^{(1)}=m_b^{(2)}=m_b^{(3)}=0$, and thus, only 
a linear combination of the three K\"ahler moduli remains massless. 
There exists one flat direction even when we consider more complicated 
configurations of the magnetic fluxes to get three or more 
unbroken gauge subgroups. 
The aim of this paper is to stabilize 
this remaining massless moduli field by nonperturbative superpotential 
originating from E-branes.

\subsection{Pati-Salam Models based on D9-branes}
\label{susec:PSmodel}
We  construct Pati-Salam models based on a stack of eight D9-branes, whose 
low-energy effective field theory is 
10D $U(8)$ SYM theory. 
In the rest of this paper, we consider 
$ Z_2\times  Z'_2$ orbifolding 
to eliminate harmful zero-modes, which acts as 
\begin{eqnarray}
 Z_2:  (z_1,z_2,z_3)&\rightarrow& (-z_1,-z_2,z_3),\nonumber\\
 Z'_2: (z_1,z_2,z_3)&\rightarrow& (z_1,-z_2,-z_3). \label{eq:z2z2}
\end{eqnarray}
Under these $ Z_2$ and $ Z'_2$ symmetries, 
the superfields transform properly (see, Eq.~(\ref{eq:Z_2transform})) 
with projection operators $P$ and $P'$, respectively. 
For later convenience 
we define the following matrix 
\begin{equation}
P_{\alpha\beta\gamma}=\begin{pmatrix}
\alpha\times {\bf 1}_4 & 0 &0\\
0 & \beta\times {\bf 1}_2 &0\\
0 & 0 & \gamma\times {\bf 1}_2
\end{pmatrix}, \label{eq:pabc}
\end{equation}
where $\alpha,\beta$ and $\gamma$ take $+1$ or $-1$ and 
${\bf 1}_n$ denotes $(n \times n)$ unit matrix. 
Orbifolding with projection operator of this form must 
respect the Pati-Salam gauge group.

In the $U(8)$ SYM theories, magnetic fluxes are represented by
three $8\times 8$ matrices. 
It is convenient to parameterize them as,
\begin{eqnarray*}
M^{(1)}&=&{\rm diag}\,(0,0,0,0,X,X,-Y,-Y)+a\times {\bm 1}_8,\\
M^{(2)}&=&{\rm diag}\,(0,0,0,0,-1,-1,0,0)+b\times {\bm 1}_8,\\
M^{(3)}&=&{\rm diag}\,(0,0,0,0,0,0,1,1)+c\times {\bm 1}_8, 
\end{eqnarray*}
%\begin{equation}
%\begin{split}
%&M^{(1)}=\begin{pmatrix}
%a\times{\bf 1}_4 & 0 &0\\
%0 & (a+X)\times{\bf 1}_2 &0\\
%0 & 0 & (a-Y)\times{\bf 1}_2
%\end{pmatrix},\\
%&M^{(2)}=\begin{pmatrix}
%b\times{\bf 1}_4 & 0 &0\\
%0 & (b-1)\times{\bf 1}_2 &0\\
%0 & 0 & b\times{\bf 1}_2
%\end{pmatrix},\ \
%M^{(3)}=\begin{pmatrix}
%c\times{\bf 1}_4 & 0 &0\\
%0 & c\times{\bf 1}_2 &0\\
%0 & 0 & (c+1)\times{\bf 1}_2
%\end{pmatrix},
%\end{split}
%\label{eq:magnetic_flux}
%\end{equation}
where $a,b,c\in\mathbb{Z}$ and $X,Y\in\mathbb{N}$. 
Note that, the 4D effective theories are independent of $a,~b$ and $c$ 
within the D9-brane sector 
except for the FI-parameters. 
They will play a significant role in association with E-branes. 
These magnetic fluxes break the $U(8)$ gauge group down to 
the Pati-Salam gauge group, $U(4)_C\times U(2)_L\times U(2)_R$ up to $U(1)$ factors, 
and produce the FI-terms for 
diagonal parts of them as 
\begin{eqnarray*}
\xi_C&=&\frac{1}{\mathcal{A}^{(1)}}a +
\frac{1}{\mathcal{A}^{(2)}}b 
+\frac{1}{\mathcal{A}^{(3)}}c,
\label{eq:fiC}\\
\xi_L&=&\frac{1}{\mathcal{A}^{(1)}}(a+X) +
\frac{1}{\mathcal{A}^{(2)}}(b-1) 
+\frac{1}{\mathcal{A}^{(3)}}c,
\label{eq:fiL}\\
\xi_R&=&\frac{1}{\mathcal{A}^{(1)}}(a-Y) +
\frac{1}{\mathcal{A}^{(2)}}b 
+\frac{1}{\mathcal{A}^{(3)}}(c+1). 
\label{eq:fiR}
\end{eqnarray*}
These FI-parameters vanish when 
\begin{equation}
\mathcal{A}^{(1)}/\mathcal{A}^{(2)}=X,\qquad 
\mathcal{A}^{(1)}/\mathcal{A}^{(3)}=Y,\qquad 
a+Xb+Yc=0.
\label{eq:D_cond}
\end{equation}
At supersymmetric vacua with the Pati-Salam gauge group, 
this implies that two of the three K\"ahler moduli 
are stabilized by the D-term.

On this magnetized orbifold with $P'=P_{+--}$ (see, Eq.~(\ref{eq:pabc})), 
there remain the following zero-modes, 
\begin{equation*}
\Phi_1=\begin{pmatrix}
0 & 0 &0\\
0 & 0 &H\\
0 & 0 & 0
\end{pmatrix},\quad 
\Phi_2=\begin{pmatrix}
0 & Q_{L}& 0\\
0 & 0 & 0\\
0 & 0 & 0
\end{pmatrix},\quad 
\Phi_1=\begin{pmatrix}
0 & 0 & 0\\
0 & 0 & 0\\
Q_R & 0 & 0
\end{pmatrix}, 
\end{equation*}
where three rows and columns correspond to 
$U(4)_C$, $U(2)_L$ and $U(2)_R$. 
We can find degenerate zero-modes in bifundamental representation 
$(1,{\bm 2},\bar{\bm 2})$, 
$({\bm 4},\bar{\bm 2},1)$ and 
$(\bar{\bm 4},1,{\bm 2})$, 
which can be identified with the Higgs fields $H$, 
the left-handed matter fields $Q_L$ and right-handed matter fields $Q_R$, respectively. 
Their degeneracy, that is, the number of generations, 
is determined by $X$, $Y$ and $Z_2$ projection operator $P$. 
Three-generation magnetized orbifold models based on 
the Pati-Salam gauge group were systematically 
studied in Ref.~\cite{Abe:2008sx}. 
According to that, we summarize 
all possible ans\"atze of $(X,\,Y,\,P)$ for realizing 
the three generations of the quarks 
and the leptons in Table \ref{tab:numH}.
\begin{table}[th]
\centering
\begin{tabular}{ccccc}
\hline
& $X$ & $Y$ & $P$ & \# of Higgs
\\
\hline
No.1 & 4 & 4 & $P_{+-+}$ & 5\\
No.2 & 5 & 5 & $P_{+-+}$ & 6\\
No.3 & 7 & 7 & $P_{++-}$ & 8\\
No.4 & 8 & 8 & $P_{++-}$ & 9\\
No.5 & 4 & 5 & $P_{+-+}$ & 5\\
No.6 & 7 & 8 & $P_{++-}$ & 8\\
No.7 & 4 & 7 & $P_{+--}$ & 5\\
No.8 & 4 & 8 & $P_{+--}$ & 5\\
No.9 & 5 & 7 & $P_{+--}$ & 5\\
No.10 & 5 & 8 & $P_{+--}$ & 6\\
\hline
\end{tabular}
\caption{They are all possible sets of $X$, $Y$ and $P$ 
to realize the three-generation structure of the SM. 
One can exchange the values of $X$ and $Y$ in configuration 5 and 6. 
In configuration 7-10, we have to replace the projection operator 
by $P_{+++}$ when exchanging $X$ and $Y$. }
\label{tab:numH}
\end{table}
In these models, 
a reasonable mechanism to realize hierarchical masses and 
mixing angles works, which leads to 
a semi-realistic spectrum without fine tunnings 
for parameters~\cite{Abe:2014vza}. 
Note that, there are other configurations to realize the three generations, 
but they have a phenomenological difficulty in textures of Yukawa matrices 
and we have omitted them here. 
It is remarkable that zero-modes cannot remain in diagonal entries of the above matrices which correspond to open string moduli fields. That is, open string moduli are completely stabilized. The idea of this open string moduli stabilization would be a T-dual picture to intersecting D-branes wrapping rigid cycles \cite{Blumenhagen:2005tn}. 

One may expect that 
magnetized backgrounds with more complicated gauge symmetry breaking, 
e.g., $U(8)\rightarrow U(3)_C \times U(1)_\ell \times U(2)_L \times U(2)_R$, 
lead to a new class of three-generation models. 
In that case, however, a nonvanishing FI-term inevitably appears 
within the 10D $U(8)$ SYM theories~\cite{Abe:2014vza}. 
We will propose such a model with all the vanishing FI parameters 
on the basis of D7-brane systems in section \ref{sec:D7-brane}.

\subsection{Nonperturbative Superpotential : E1-branes}
\label{sec:e1d9}
We study E-branes in D9-brane models.
In general, there can be various E-branes generating superpotential.
Here, we focus only on E-brane configurations which contribute to the moduli stabilization. 
%nonperturbative superpotential 
%for the moduli fields. 
In the presence of D9-branes, 
two types of E-branes are possible to lead to a stable brane system; 
E1-branes wrapping two-cycles 
and E5-branes. 
E-branes generically have $O(N)$ or $USp(N)$ gauge groups, and 
only the $O(N)$-type instantons can generate the superpotential. 
In the present setup, we can choose discrete torsions to obtain the $O(N)$-type instantons \cite{Blumenhagen:2005tn}, 
and we assume that the discreet torsions are tuned on suitably in this paper. 

These instantons can induce superpotential of the form 
\begin{equation}
W_{np}=\sum_{i} A_i e^{-a_i T_i} +A_Se^{-S},\label{eq:supapo}
\end{equation}
where $T_i$ and $S$ are K\"ahler moduli and dilaton superfields, respectively. 
Coefficients $A_i$ and $A_S$ depend on complex moduli fields, 
which are supposed to be stabilized by the 3-form fluxes and replaced by their VEVs. 
In the present case, they are given by 
\begin{equation*}
T_i=e^{-\phi} {\mathcal A}^{(i)}+i\int_{T^{2}} C_2,\qquad 
S=e^{-\phi} {\mathcal A}^{(1)}{\mathcal A}^{(2)}{\mathcal A}^{(3)}
+i\int_{T^6}C_6, 
\end{equation*}
where $C_2$ and $C_6$ are RR-forms and $\phi$ is the 10D dilaton field. 
The SUSY condition (\ref{eq:D_cond}) stabilizes 
two directions of $T_i$. 
It is important that 
this superpotential changes or vanishes if there exist open string zero-modes 
between the D9-branes and the E-branes.  
We have to study configurations of these branes 
in order to eliminate such harmful zero-modes.

First we study E1-branes, which wrap one of the three $T^2$ and 
are collapsed at a fixed point on the other $T^2$. 
A single E1-brane has an $O(1)$ gauge symmetry and is 
to generate the superpotential for the K\"ahler moduli 
(the first term of Eq.~(\ref{eq:supapo})) 
as long as there is no extra zero-mode. 
A zero-mode configuration of E1/D9 systems is equivalent to that of a system 
consisting of 
D9-branes and an unfluxed D5-brane wrapping 
the $i$-th $T^2$. 
Such a D-brane system contains a six-dimensional $\mathcal N=1$ 
hypermultiplet as D5-D9 (or E1-D9) open strings. 
Naming these D9-branes ``D9$_A$'', 
we can represent the hypermultiplet 
by using two 4D $\mathcal N=1$ superfields 
as ($\Phi^{AE}_j, \Phi^{EA}_k$) ($i\neq j\neq k\neq i$) in 
the superfield description (see, Ref.~\cite{Abe:2015jqa}). 
Note that superscripts $AE$ and $EA$ reflect their gauge transformation laws, 
and they are (anti-)fundamental representation of $U(N)$ gauge group of 
the D9-branes. 
They are affected by the magnetic fluxes of the D9-branes, 
and thus a chiral spectrum with generation structure is 
produced in this E1-D9 sector, 
in the same way as D9-brane sector. 
The transformation law of these chiral superfields 
under the $Z_2$ and $Z_2'$ orbifolding 
is given in a way similar to the D9-brane fields, e.g., 
\begin{equation*}
\Phi^{AE}_1 \rightarrow - P \Phi^{AE}_1 P_E^{-1},\qquad 
\Phi^{AE}_1 \rightarrow + P' \Phi^{AE}_1 P_E^{'-1}, 
\end{equation*} 
where we can set $P_E$ and $P_E'$ to $\pm1$. 
Note that all of E1-branes with $P_E = \pm 1$ and $P_E'=\pm1$ can appear and 
we have to take into account all the possible E1-branes including projections,  $P_E = \pm 1$ and $P_E'=\pm1$.
However, some of them do not induce nonperturbative terms and others induce nonperturbative terms 
such as (\ref{eq:supapo}) as well as nonperturbative terms with matter fields.
We are interested in E1-branes with proper orbifold parities, $P_E$ and $P_E'$, which can induce 
(\ref{eq:supapo}).
When the superfield has  a different subscript, 
the overall signs can be changed. 
Their wavefunctions can be even or odd functions 
on the $i$-th $T^2$. 
On the other $T^2$, however, 
they cannot survive the orbifold projection when they are 
assigned into odd mode, 
because they are localized at a fixed point of the $T^2$ and 
their wavefunctions must be given by a delta function.

We study how to find the E1-brane configurations where 
all the harmful 
massless modes are eliminated, taking 
an E1-brane wrapping the third $T^2$ as an example.
For the purpose, it is satisfactory to 
investigate a zero-mode configuration of 
$\Phi^{AE}_1$ and $\Phi^{EA}_2$. 
They transform under the $Z_2$ symmetry as 
\begin{equation*}
\Phi^{AE}_1 \rightarrow - P \Phi^{AE}_1 P_E^{-1},\qquad 
\Phi^{EA}_2 \rightarrow - P_E\Phi^{EA}_2P^{-1}. 
\end{equation*} 
They cannot have zero-modes 
when they are assigned into $Z_2$ odd mode 
on the first and the second $T^2$ as discussed above. 
Thus, for $P=P_{+++}$, we can eliminate 
all the components of $\Phi^{AE}_1$ and $\Phi^{EA}_2$ by 
$P_E=+1$. 
Even when $P\neq P_{+++}$, it is possible to eliminate them as follows. 
In the Pati-Salam models,  both of them have eight components, which are 
classified into three parts by their gauge representations, 
i.e., $U(4)_C$, $U(2)_L$ and $U(2)_R$. 
A proper choice for $P_E$ can forbid the charged zero-modes 
in two of the three parts. 
Seen from Table~\ref{tb:numzero}, we can eliminate the remaining ones 
when the absolute values of their effective magnetic fluxes are 
less than three and they are assigned into $Z_2'$ odd mode on the 
third $T^2$. 
We can always find $P_E'$ and $c$ which realize such a situation, 
satisfying Eq.~(\ref{eq:D_cond}). 
Thus, it is always possible for the E1-brane 
to generate the nonperturbative superpotential. 
One can easily confirm that E1-branes wrapping the first or 
the second $T^2$ can also induce nonperturbative terms to stabilize the moduli.

We examine the stabilization of the moduli field 
minimizing its potential. 
We expect to obtain the following nonperturbative 
superpotential, 
\begin{equation} 
W=Ae^{-2\pi T_3}+W_0.\label{eq:tstabi} 
\end{equation}
We assumed that nonperturbative term due to E1-brane wrapping the third $T^2$ is dominant.
Even when other terms are dominant, the following discussion is the same.
A constant term $W_0$ is also necessary for the moduli stabilization, 
and we will discuss its origin later. 
In toroidal compactifications of type IIB with O5/O9 planes, 
the K\"ahler potential for the moduli fields 
is given by 
\begin{equation*}
K_0=-\log(S+\bar S)-\sum_{i=1}^3\log(T_i+\bar{T_i}) 
-\sum_{i=1}^3\log(U_i+\bar{U_i}). 
\end{equation*}
Setting ${\rm Re\,}T_i=\tau_i$ and ${\rm Im\,}T_i=0$, 
we get the F-term potential 
\begin{equation*}
V_F=\frac{\pi Ae^{-4\pi\tau_3}}{\tau_1 \tau_2}\left(A+2\pi A\tau_3+W_0e^{2\pi \tau_3}\right).
\end{equation*}
Minimizing this potential, we find a supersymmetric minimum,
\begin{equation*}
\frac{W_0}A=-(1+4\pi\tau_3)e^{-2\pi \tau_3}, 
\end{equation*}
where $\tau_3$ is stabilized. 
In this case, one sees that 
a legitimate value of $\langle\tau_3\rangle$ implies 
a quite small value of $W_0$, indeed, 
$\langle\tau_3\rangle=1$ requires $W_0/A\sim 10^{-2}$. 
The origin of such a small $W_0$ will be discussed in the next subsection.

\subsection{Nonperturbative Superpotential : E5-branes}
\label{sec:E5}

We  perform a study similar to the previous subsection for E5-branes. 
The number of zero-modes in D9-E5 open strings 
can be counted in the same way as a mixed configuration of 
the magnetized D9-branes and an additional D9-brane with no magnetic fluxes.

Although it is difficult in D9/E5 systems 
to give a setup to eliminate all the harmful zero-modes systematically, 
we show a reasonable setup to generate the nonperturbative superpotential 
to be incorporated in a wide class of the Pati-Salam models shown 
in Table~\ref{tab:numH}. 
First we set $b=-1$ and $c=+1$, which implies $a=X-Y$ 
for the vanishing D-terms (see, Eq.~(\ref{eq:D_cond})). 
That is, the magnetic fluxes in the Pati-Salam sector are given by 
\begin{eqnarray*}
M^{(1)}&=&{\rm diag}\,(X-Y,X-Y,X-Y,X-Y,2X-Y,2X-Y,X-2Y,X-2Y),\\
M^{(2)}&=&{\rm diag}\,(-1,-1,-1,-1,-2,-2,-1,-1),\\
M^{(3)}&=&{\rm diag}\,(1,1,1,1,1,1,2,2).
\end{eqnarray*}
When $2X-Y\neq0$ and $X-2Y\neq0$, 
an association of chirality projections due to the magnetic fluxes 
and $Z_2'$ orbifold projections with $P_E'=-1$ eliminates 
open string zero-modes charged under $U(2)_L$ and $U(2)_R$. 
The remaining ones, which are 
(anti-)fundamentals in the $U(4)_C$ gauge group, 
can also be eliminated by $Z_2$ orbifolding with a suitable choice for $P_E$, 
if $0<|X-Y|<3$. 
Thus we can always provide configurations of D9/E5 systems 
to generate 
the nonperturbative superpotential for the dilaton superfield 
(the second term of Eq.~(\ref{eq:supapo})), 
when $X$ and $Y$ satisfy 
\begin{equation*}
2X-Y\neq0,\qquad X-2Y\neq0,\qquad 0<|X-Y|<3. 
\end{equation*}
That is, models 5,\,6 and 9 shown in Table~\ref{tab:numH} 
are available (We can exchange the values of $X$ and $Y$ as discussed there). 
In particular, 
we find that some of these models can be associated with 
an E5-brane and an E1-brane simultaneously 
(e.g., $X=7$, $Y=5$ and $P=P_{+++}$). 
In this case, 
we can obtain the nonperturbative superpotential
\begin{equation}
W=A_Ee^{-2\pi T_3}+A_Se^{-S}.\label{eq:wst}
\end{equation}
We have assumed the presence of supersymmetric 3-form fluxes to 
stabilize the dilaton satisfying $\langle W_{\rm 3-form}\rangle=0$. 
The superfield $S$ can be replaced by its VEV, and then 
the effective superpotential is equivalent to Eq.~(\ref{eq:tstabi}), 
that is, 
\begin{equation*}
W_0=A_Se^{-\langle S\rangle}. 
\end{equation*}
From this expression, it is found that 
a reasonable value of $\langle S\rangle$ induces a 
sufficiently small $W_0$ which is required for the above 
successful moduli stabilization. 
Thus, all the moduli fields can be stabilized 
in the framework of magnetized D-branes by 
an interplay of the two instanton effects.

\section{D7-brane models}
\label{sec:D7-brane}

In this section, we consider another model based on D7-branes, instead of 
the Pati-Salam models based on D9-branes\footnote{
The model discussed in this section was proposed in Ref.~\cite{horie}}. 
\subsection{MSSM-like model}
We consider an MSSM-like model on the basis of 
two stacks of four D7-branes which we denote by 
D7$_A$-branes and D7$_B$-branes with a configuration 
shown in Table \ref{tab:D7system}. 
\begin{table}[t]
\begin{center}
\begin{tabular}{cccc}
 &$T^2$ &$T^2$&$T^2$\\\hline
${\rm D7}_A$& \checkmark & $\times$ & \checkmark \\
${\rm D7}_B$& \checkmark & \checkmark & $\times$ 
\end{tabular}
\end{center}\vspace{-1.5em}
\caption{The configuration of two stacks of D7-branes is shown. 
A symbol ``\checkmark'' means that D-branes wrap $T^2$, and 
another one ``$\times$'' expresses that 
D-branes are localized at a fixed point on $T^2$. }\label{tab:D7system}
\end{table}
An effective field theory of D7-branes 
is derived from a 10D SYM theory, 
and the superfield description of that was formulated 
in Ref.~\cite{Abe:2015jqa}. 
One of the three chiral superfields $\Phi_i$ contained in 10D SYM theories 
turns to a position moduli field there. 
In the present case of the mixed D7-brane system, 
there also appears a hyper multiplet corresponding to 
open string modes between the D7$_A$- and D7$_B$-branes, 
which is denoted by two chiral superfields $\Phi_2^{AB}$ and $\Phi_3^{BA}$. 
Thus, this system consists of the following chiral superfields, 
\begin{equation*}
\Phi_1^A,\quad \tilde\Phi_2^A,\quad \Phi_3^A,\quad \Phi_1^B,
\quad \Phi_2^B,\quad \tilde\Phi_3^B,\quad \Phi_2^{AB},\quad 
\Phi_3^{BA}. 
\end{equation*}
The first three superfields are in the $U(4)_A$ adjoint representation, 
and the next three are in the $U(4)_B$ adjoint one. 
The last two are bifundamental representation of $U(4)_A\times U(4)_B$. 
The tilde represents that the superfield turns to be a position moduli 
of the corresponding D7-branes. 

In this section, we again consider $T^2\times T^2\times T^2$ as the 
extra compact space with $Z_2\times Z_2'$ orbifolding. 
These $Z_2\times Z_2'$ act on the three $T^2$ in the same way as 
in the previous section, and 
the transformation laws of the superfields are determined by 
their subscript and 
four $4\times 4$ projection matrices, $P_A$, $P_B$, $P'_A$ 
and $P'_B$. 
Note that, active D7-brane fields must be assigned into even mode on 
$T^2$ where the D7-brane is localized as a point because 
such a point-like localization implies a wavefunction of delta function. 
In particular, 
D7-D7 open strings, $\Phi_2^{AB}$ and $\Phi_3^{BA}$, 
have to be assigned into even mode on the second and the third $T^2$ 
in order to survive the orbifold projections.

We introduce the magnetic fluxes 
in this D7$_A$/D7$_B$ brane system 
as follows,
\begin{equation*}
M_A^{(1)}=\begin{pmatrix}
-5 \times{\bf 1}_3 & 0 \\
0 & -4\times{\bf 1}_1 
\end{pmatrix},\qquad 
M_A^{(3)}=\begin{pmatrix}
5\times{\bf 1}_3 & 0 \\
0 & 4\times{\bf 1}_1 
\end{pmatrix},
\label{eq:E7-1}
\end{equation*}
\begin{equation*}
M_B^{(1)}=\begin{pmatrix}
0 \times{\bf 1}_3 & 0 \\
0 & -12\times{\bf 1}_2 
\end{pmatrix},\qquad 
M_B^{(2)}=\begin{pmatrix}
0 \times{\bf 1}_2 & 0 \\
0 & 1\times{\bf 1}_2 
\end{pmatrix}.
\label{eq:E7-2}
\end{equation*}
These magnetic fluxes break 
$U(4)_A\times U(4)_B \rightarrow U(3)_C\times U(1)_l \times U(2)_L \times U(2)_R$. 
One remarkable feature of this model is breaking of the $U(4)_C$ 
gauge symmetry of the Pati-Salam models. 
This means that the quarks and the leptons 
can have a distinguished difference in their flavor structure. 
The flux-induced FI-terms vanish in all the 
unbroken gauge subgroups when 
\begin{equation}
\mathcal{A}^{(1)}/\mathcal{A}^{(2)}=12\quad{\rm and}\quad 
\mathcal{A}^{(1)}/\mathcal{A}^{(3)}=1. \label{eq:d7fi}
\end{equation}
Setting the projection operators as $P_A=P_B=P_A'={\bm 1}_4$ 
and $P_B'=-{\bm 1}_4$, we find the following zero-mode structure, 
\begin{equation*}
\Phi_1^B=\begin{pmatrix}
 0 &H\\
 0 & 0
\end{pmatrix},~~
\Phi_2^{AB}=\begin{pmatrix}
Q_L & 0\\
L_L & 0
\end{pmatrix},~~
\Phi_3^{BA}=\begin{pmatrix}
0 & 0 \\
Q_R & L_R 
\end{pmatrix},~~
\end{equation*} 
and $\Phi_1^A, \Phi_2^A, \Phi_3^A, \Phi_2^B$ and $\Phi_3^B$ have no zero-mode. 
We can identify 
$H$, $Q_L$, $Q_R$, $L_L$ and $L_R$ with 
the Higgs fields, the left-handed quarks, the right-handed quarks, 
the left-handed leptons and the right-handed leptons of the MSSM, 
respectively. 
All of the position and Wilson-line moduli fields are stabilized in this model as well as in the D9-models. 

\subsection{Nonperturbative Superpotential : E3-branes}

In the present D7-brane system, 
there are two types of E-branes keeping the whole brane system stable; 
E3-branes and E(-1)-branes. 
When there are no open string zero-modes interplaying 
the D-branes and the E-branes, 
these instantons generate the nonperturbative 
superpotential (Note again that, we have assumed discrete torsions tuned on to obtain 
$O(N)$-type E-branes.), 
\begin{equation}
W_{{\rm np}}=\sum_{i} A_i e^{-a_i T_i} +A_S e^{-S}. \label{eq:d7super}
\end{equation}
In the IIB orientifold with $O3/O7$-planes, $T_i$ and $S$ are given by, 
($i\neq j\neq k\neq i$)
\begin{equation*}
T_i=e^{-\phi} {\mathcal A^{(j)}}{\mathcal A^{(k)}}+ i\int_{T^4}C_4, 
\qquad S=e^{-\phi} + i C_0, 
\end{equation*}
and again, we see that two of the K\"ahler moduli fields are stabilized 
by Eq.~(\ref{eq:d7fi}). 

We first discuss an E3-brane wrapping two of three $T^2$ 
and localized at a fixed point on the other one, 
which has an $O(1)$ gauge symmetry and generates the first term 
superpotential of Eq.~(\ref{eq:d7super}), without extra zero-modes. 
Generic E3/D7 systems are classified into two cases.
One is the case 
when the E-branes and the D-branes wrap the same $T^4=T^2\times T^2$ and 
localized at fixed points on the last $T^2$. In this case, 
it is easy to eliminate E3-D7 open string zero-modes, 
because the two stacks of the branes can be sequestered spatially when 
the two stacks are localized at different fixed points. 
In the other case, 
we have to study the zero-mode distribution in detail for each model. 
Recall again that any E3-brane including all the possible positions and orbifold parities 
can appear and we have to take into account all the possibilities.
However, we are interested only in E3-brane configurations to lead moduli-dependent superpotential terms.

In the present D7-brane system, 
there are two stacks of D7-branes which wrap the different four directions 
of extra compact space. 
An additional E-brane can be sequestered from one stack by 
a localization at different fixed points, 
but there exist 
massless open strings between the E-brane and the other stack of D7-branes 
to be eliminated by the orbifold projection. 
Let us consider an E3-brane which wraps the first and the second $T^2$ 
and is localized at a ``vacant'' fixed point on the third $T^2$. 
E3-D7$_B$ zero-modes cannot appear, but 
there are E3-D7$_A$ open strings denoted by $\Phi_2^{AE}$ and $\Phi_3^{EA}$. 
Fortunately, we can eliminate them easily as follows. 
They transform under the $Z_2'$ symmetry ($P_A'=+{\bm 1}_4$) as 
\begin{equation*}
\Phi_2^{AE}\rightarrow -\Phi_2^{AE}P_E',\qquad 
\Phi_3^{EA}\rightarrow -P_E^{'-1}\Phi_3. 
\end{equation*}
Thus they all can be assigned into $Z_2'$ odd mode by $P_E'=+1$ 
and are eliminated as we wanted, 
because wavefunctions of these open strings must be 
an even function on the second and the third $T^2$ 
as explained above. 
As a result we obtain the superpotential 
\begin{equation*}
W=A_3e^{-2\pi T_3}. 
\end{equation*}
We will see that 
this stabilizes the moduli in association 
with an additional E(-1)-brane in the following subsection. 

\subsection{Nonperturbative Superpotential : E(-1)-branes}
It is much easier to find an E(-1)-brane configuration generating 
the superpotential for the dilaton superfield.
The E(-1)-brane is an instanton localized completely 
at a point on the whole compact space. 
Thus we can trivially sequester the E(-1)-brane from the D7-brane 
system in order not to produce the harmful zero-modes, 
unless four fixed points of a $T^2/Z_2(Z'_2)$ are 
occupied by multiple stacks of D7-branes. 

One can straightforwardly see that the D7-brane system admits 
an E(-1)-brane and an E3-brane simultaneously 
%in the D7-brane model obtain 
and superpotential~(\ref{eq:wst}) is generated. 
Similarly to the D9-brane systems, 
the second term of Eq.~(\ref{eq:wst}) produces a small constant term 
in the superpotential, 
and the K\"ahler moduli field is stabilized 
with a moderate value of the VEV.

\section{Conclusion and Discussion} 
\label{sec:con}
We have studied the nonperturbative superpotential induced 
by E-branes in semi-realistic D-brane models based on 
the toroidal orbifolds. 

We have considered two types of D-brane models for the visible sector. 
One is based on a stack of eight D9-branes, where 
the magnetic fluxes and the orbifold projection yield the 
Pati-Salam gauge group with the three generations of the quarks and 
the leptons. 
In the models, magnetic fluxes generate FI-terms, which depend on the K\"ahler moduli, 
and those fix the ratio among three K\"ahler moduli.
Furthermore, we have found that 
an E1-brane and an E5-brane generate the superpotential for the dilaton 
and the K\"ahler moduli, respectively. 
The dilaton is replaced by its VEV in the nonperturbative superpotential 
because we have assumed the 3-form fluxes to stabilize that. 
That gives rise to the sufficient small constant term, and as a result, 
the K\"ahler moduli field is stabilized with 
a moderately large value of the VEV. 
The other D-brane model is derived from 
the two stacks of the D7-branes. 
In this model, the moduli-dependent FI-terms can fix the ratio of three K\"ahler moduli.
On top of that, we have found 
that an E3-brane and an E(-1)-brane successfully generate 
the superpotential and stabilize the moduli. 
In our study, we have found 
some constraints on the magnetic fluxes and the orbifold parities 
for realizing the moduli stabilization, and 
it is quite nontrivial that there exists 
a successful configuration of D-branes and E-branes.

%These magnetized D-brane models are T-dual to intersecting D-branes. 
%The T-dual picture allows us to count the number of the remaining zero-modes 
%in a simple formula. 
%In the T-dual picture, we have indeed regained  
%the consequence obtained in the magnetized D-brane models. 

In this paper, 
we have studied moduli stabilizations with only the visible sector. 
The vacuum is the supersymmetric vacuum with negative energy.
We need SUSY breaking and uplifting the vacuum energy to almost zero energy.
Thus, towards more realistic models, we should also 
consider a hidden sector for SUSY breaking.\footnote{See e.g. for explicit construction of the SUSY breaking 
sector \cite{Abe:2016zgq}.} 
In that case, we have to care about open string zero-modes 
between the E-branes and the hidden D-branes, 
because the moduli stabilizing superpotential vanishes if 
there appears an extra zero-mode. 
Besides the open string zero-modes, we expect that 
there are several important interplays between 
the SUSY breaking and the moduli stabilization. 
It seems that such an extension to contain the SUSY breaking sector 
is a very challenging task towards realistic D-brane models. 

Nonperturbative effects due to E-branes are applied to other 
phenomenological issues than the moduli stabilization. 
Another challenging task of D-brane models is 
to obtain Majorana mass terms 
and supersymmetric Higgs mass term ($\mu$-term). 
We are able to consider additional E-branes to generate 
these mass terms~\cite{Blumenhagen:2006xt,Ibanez:2006da,Ibanez:2007rs,Cvetic:2007ku,Blumenhagen:2009qh,Kobayashi:2015siy,Kobayashi:2016ovu}. 
It is also an attractive prospect to try that 
in the D-brane models shown in this paper. 

\section*{Acknowledgments}

H.~A. was supported by JSPS KAKENHI Grant Number JP16K05330. 
T.~K. was supported in part by JSPS KAKENHI Grant Number JP26247042. 
K.~S. was supported by Waseda University Grant for Special Research Projects No.~2016B-200. 
S.~U. was supported by JSPS KAKENHI Grant Number JP15J02107. 

\appendix

\section{T-dual picture} 
\label{sec:T-dual}

Magnetized D-brane systems are T-dual to intersecting D-branes. 
Although they are physically equivalent to each other, 
one may easily be able to investigate the remaining zero-modes 
in intersecting D-brane systems 
than in magnetized D-brane systems. 
In this appendix, we introduce an instrument 
to count the active zero-modes in the T-dual picture, 
i.e. intersecting D6-branes 
wrapping rigid 3-cycles on $T^6/ Z_2\times  Z_2'$ with discrete torsion 
(See Ref.~\cite{Blumenhagen:2005tn} for reference.).

\subsection{T-dual to D9-brane models} 
\begin{figure}
  \centering
  \includegraphics[width=12cm]{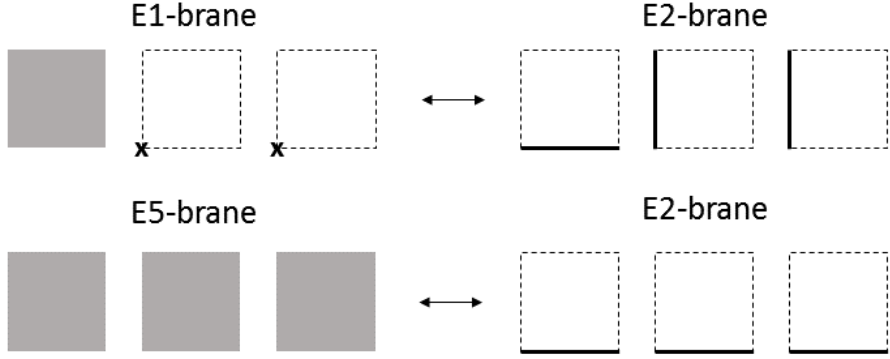}
  \caption{A set of three squares expresses 
a fundamental region of $T^2\times T^2\times T^2$. 
The left and right sides correspond to each other 
in T-duality along three vertical axes. }\label{fig:tdual}
\end{figure}

E1-branes and E5-branes discussed in section \ref{sec:d9} 
are both equivalent to E2-branes wrapping 
different extra dimensional directions in the T-dual side. 
That is schematically depicted in Fig.~\ref{fig:tdual}. 
Note that, this figure does not take into account orbifolding 
for simplicity.

The number of zero-modes between an E2-brane and a D6-brane 
is counted by the topological intersection number \cite{Blumenhagen:2005tn}, 
\begin{equation}
\frac{1}{4} \prod_{i=1}^3\left(n_E^{(i)}\tilde m_D^{(i)} 
-\tilde m_E^{(i)} n_D^{(i)} \right)+\frac{1}{4}\sum_{g\in G} \sum_{i,j\in J_g^E} \sum_{k,l \in J^D_g} \epsilon_{E,ij}^g \epsilon_{D,kl}^g \delta_{ik}\delta_{jl} (n_E^{(I_{g})} \tilde{m}_D^{(I_g)} - \tilde{m}_E^{(I_g)} n_D^{(I_g)}). 
\label{eq:T-dualnu}
\end{equation}
In this expression, subscripts $E$ and $D$ express the E-brane 
and the D-brane. When we denote nontrivial elements of $Z_2$ and $Z_2'$ 
by $\theta$ and $\theta'$, respectively, $G$ is a set of 
$\theta$, $\theta'$ and $\theta\theta'$. 
For each $g$, 
sets of fixed points where the E-branes and the D-branes live 
are given by $J_E^g$ and $J_D^g$, respectively. 
There are two possible orientations on 
2-cycles collapsed at a fixed point contained in $J_E^g$ or $J_D^g$. 
This degree of freedom is defined by $\epsilon_{a,ij}^g=\pm 1$. 
In the magnetized D9-brane models, 
that corresponds to the discrete Wilson lines and parities 
$P$ and $P'$ 
(We have not considered the Wilson lines in this paper, 
and then we get $\epsilon_{E,ij}^g=\epsilon_{E}^g$ 
and $\epsilon_{D,kl}^g=\epsilon_{D}^g$.). 
A set of $(n_D^{(i)} ,m_D^{(i)})$ represents winding numbers along 
two fundamental cycles of the $i$-th $T^2$, 
and $(n_a^{(I_g)},m_a^{(I_g)})$ denotes winding numbers 
on a $T^2$ invariant under $g\in  Z_2\times  Z_2'$. 
That is, in the present case (\ref{eq:z2z2}), 
we see $(I_\theta,I_{\theta'},I_{\theta\theta'})=(3,1,2)$. 
The tilde on the winding number is a reflection of 
nontrivial complex structure, e.g., 
$\tilde m^{(i)}_D\equiv m^{(i)}_D+\frac{1}{2}n_D^{(i)}$ when the torus is tilted, and $\tilde m^{(i)}_D= m^{(i)}_D$ when the torus is rectangular. 
In the following, we take $\tilde m^{(i)}=m^{(i)}$ for simplicity 
which is satisfactory for the aim of this section.

In the upper case of Fig.~\ref{fig:tdual}, 
the winding numbers of the corresponding E2-brane are given by 
\begin{equation}
(n_E^{(1)},m_E^{(1)})=(-1,0),\qquad (n_E^{(2)},m_E^{(2)})=(0,1),\qquad (n_E^{(3)},m_E^{(3)})=(0,1). 
\label{eq:E1-brane}
\end{equation}
In the lower case, the winding numbers are 
\begin{equation}
(n_E^{(i)},m_E^{(i)})=(1,0)\qquad\forall i. \label{eq:e5tdual}
\end{equation}

For example, we will count the number of 
E2-D6 open string zero-modes with winding numbers (\ref{eq:E1-brane}). 
The corresponding intersection number is given by
\begin{equation*}
I_{AE}= \frac{1}{4} m_D^{(1)} +\frac{1}{4} \epsilon_{D}^{\theta'} \epsilon_{E}^{\theta'}  m_D^{(1)} +\frac{1}{4}  \sum_{i,j\in S_{\theta'}^a} \sum_{k,l \in S_{\theta'}^b} \epsilon_{D}^{\theta} \epsilon_{E}^{\theta} \delta_{ik}\delta_{jl}(1+\epsilon_{D}^{\theta'} \epsilon_{E}^{\theta'}),
\end{equation*}
One see that this intersection number vanishes 
when $\epsilon_{D}^{\theta'} \epsilon_{E}^{\theta'} = -1$. 
Thus we can add an E2-brane into the Pati-Salam models based on 
the eight D6-branes to generate the moduli stabilizing superpotential, 
if $\epsilon_{D_a}^{\theta'} \epsilon_{E}^{\theta'} = -1$ is held for all of 
the eight D6-branes ($a=1,2,\ldots,8$). 
This implies 
$\epsilon_{D_1}^{\theta'}=\epsilon_{D_1}^{\theta'}=\cdots=\epsilon_{D_8}^{\theta'}$. 
The same result has been obtained in the magnetized D-brane systems, 
that is, we have shown that the zero-modes of E1/D9 open strings 
are completely eliminated for $P=P_{+++}$ in subsection~\ref{sec:e1d9}. 
Similarly one can regain the result obtained in section~\ref{sec:E5} 
by using the winding numbers (\ref{eq:e5tdual}).

\subsection{T-dual to D7-brane models}
\begin{figure}
  \centering
  \includegraphics[width=12cm]{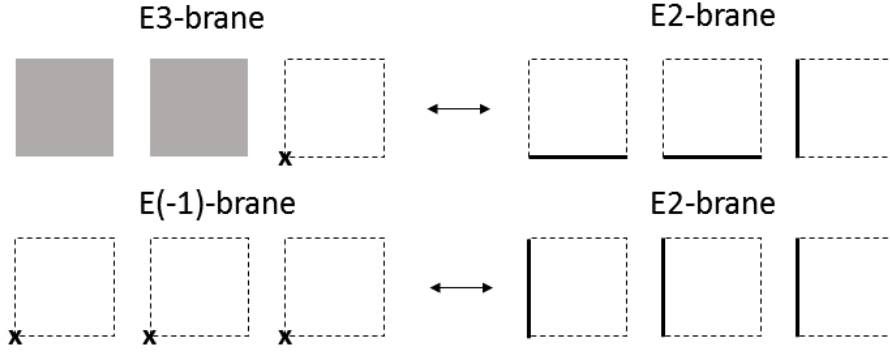}
  \caption{A set of three squares expresses 
a fundamental region of $T^2\times T^2\times T^2$. 
The left and right sides correspond to each other 
in T-duality along three vertical axes. }\label{fig:tdual2}
\end{figure}

We study the T-dual picture of the D7-brane model with 
E3- and E(-1)-branes. 
They correspond to two types of E2-branes in the T-duality 
as shown in Fig.~\ref{fig:tdual2}. 
In the upper case, 
the winding number of the E2-brane is given by 
\begin{equation*}
(n_E^{(1)},m_E^{(1)})=(1,0),\qquad 
(n_E^{(2)},m_E^{(2)})=(1,0),\qquad 
(n_E^{(3)},m_E^{(3)})=(0,1). 
\end{equation*}
In the other case one see
\begin{equation*}
(n_E^{(i)},m_E^{(i)})=(0,1). \qquad\forall i 
\end{equation*}
Substituting them in Eq.~(\ref{eq:T-dualnu}), 
one is able to confirm the result obtained in the previous section.

\end{document}